\documentclass[aps,prb,twocolumn,groupedaddress]{revtex4}
\usepackage{epsfig}
\usepackage[american]{babel}
\usepackage{amsmath}
\usepackage[dvips]{color}

\usepackage[latin2]{inputenc}
\usepackage[T1]{fontenc}
\usepackage[american]{babel}
\usepackage{color}
\usepackage{graphicx}
\usepackage{graphics}

\newcommand{\hc}{\mbox{$h. c.$}}

\begin{document}

\title{Slave fermion approach to the metallic phase in large $U_d$ high-T$_c$  cuprates}

\author{S. Bari\v si\' c}

\affiliation{Department of Physics, Faculty of Science, University of
Zagreb, Bijeni\v cka c. 32, HR-10000 Zagreb, Croatia }

\email{sbarisic@phy.hr}

\author{O. S. Bari\v si\' c}

\affiliation{Jo\v zef Stefan Institute, SI-1000 Ljubljana, Slovenia}

\affiliation{Institute of Physics, Bijeni\v cka c. 46, HR-10000 Zagreb, Croatia}

\begin{abstract}

Atomic repulsion $U_d$ on the Cu site in high T$_c$ cuprates is large but, surprisingly, some important properties are consistent with moderate couplings. The time dependent perturbation theory with slave particles is therefore formulated in the $U_d\rightarrow\infty$ limit for the metallic phase in the physically relevant regime of the three-band Emery model. The basic theory possesses the local gauge invariance asymptotically but its convergence is fast when the average occupation of the Cu-site is small. The leading orders exhibit the band narrowing and the dynamic Cu/O$_2$ charge transfer disorder. The effective local repulsion between particles on oxygen sites is shown to be moderate in the physical regime under consideration. It enhances the coherent incommensurate SDW correlations. The latter compete with the Cu/O$_2$ charge transfer disorder, in agreement with basic observations in high T$_c$ cuprates.

\end{abstract}

%\pacs{}

\maketitle

It has long been maintained \cite{fr1,fr3} and recently reemphasized \cite{an1} that one of the most important unanswered questions in the theory of the high-T$_c$ superconductivity in cuprates concerns the nature of metallic correlations which reduce the large Coulomb interaction $U_d$ on the Cu sites. In the present work we discuss this issue within the widely used large $U_d$ Emery model \cite{em1} which adds two oxygen sites O$_{x,y}$ to each Cu in the CuO$_2$ unit cell. The large $U_d$ limit is usually taken perturbatively in the Cu-O hybridization, starting from the unperturbed state with $n_d^{(0)}=1$. This leads to the superexchange \cite{ge1,za1} $J$ as the basis of the popular $tJ$ models \cite{za1}. However, various local measurements for doping $|x|\leq0.2$, NQR in particular, indicate \cite{ku1} that the average occupation of the Cu site is $n_d\approx1/2$.  We argue here that the $n_d^{(0)}=0$ unperturbed ground state is a good starting point for the large $U_d$ perturbation theory associated with the metallic phase at appreciable $|x|$, while the $x\approx0$ Mott-AF phase and the short range AF order can be approached from $n_d^{(0)}=1$ ($tJ$ model) side. Thus we generalize to finite Cu-O hybridization the early proposal \cite{go1} that in the high T$_c$ cuprates the $d$-level with large $U_d$ falls close to the nearly free (oxygen) band and that it can be related to the $d_{10}\leftrightarrow d_9$ disorder effects. In particular we show that, in the presence of the Cu-O hybridization, the $d_{10}\leftrightarrow d_9$ disorder is dynamic, while the effective interaction between the carriers propagating on oxygens is repulsive and reasonably small.

As usual \cite{em1}, the Cu,O site energies are denoted here by $\varepsilon_d$, $\varepsilon_p$ and $\Delta_{pd}=\varepsilon_p-\varepsilon_d>0$ in the hole picture. The Cu-O and O$_x$-O$_y$ hoppings are $t_{pd}$ and $t_{pp}$, the latter being retained although smaller than $t_{pd}$. Finally $1+x$ denotes the total number of holes per CuO$_2$ unit cell. The equal sharing of the hole charge between Cu and two O´s in the metallic phase is characteristic of the free fermion $t_{pd}\gg\Delta_{pd}$ model \cite{fr1,bs5}. The values $n_d\approx1/2$ were also shown \cite{mr1,su2} to be consistent with the mean-field slave boson (MFSB) fits of the ARPES results, which suggest the regime $2t_{pd}^2\leq-\Delta_{pd}t_{pp}$.   This motivates us to investigate here the $U_d=\infty$ limit of the finite $|x|$ Emery model  for $\Delta_{pd}$ sufficiently small and $t_{pd}>|t_{pp}|$.

        To this end, we use the $U_d=\infty$ slave particle theory \cite{bn1,li1,ko1,mr1}. The three-state space ($d^8$ state omitted) associated with the $\vec R$-th Cu site is described using the $f_{\vec R}^\dagger$ and $b_{\vec R,\sigma}^\dagger$ slave particles and the physical fermion $c_{\vec R,\sigma}^\dagger$ is represented as $c_{\vec R,\sigma}^\dagger\rightarrow b_{\vec R,\sigma}^\dagger f_{\vec R}$. The anticommutation rules of the physical fermions $c_{\vec R,\sigma}^\dagger$ on and among Cu sites are satisfied, provided that the $f_{\vec R}^\dagger$ and  $b_{\vec R,\sigma}^\dagger$ slave particles are taken, respectively, as fermions and bosons \cite{li1} (slave fermion theory; SFT) or vice versa \cite{bn1} (slave boson theory; SBT). In both representations the number operators satisfy $Q_R=n_{f_R}+\sum_\sigma n_{b_R,\sigma}=1$. The corresponding number operators obey $n_{d_R,\sigma}= n_{b_R,\sigma}$, which is usually called \cite{mr1} the Luttinger sum rule (LSR). The anticommutations between the Cu and O-sites are satisfied in the SBT because $b_{\vec R,\sigma}^\dagger$ and $p_{\vec R_0,\sigma'}^\dagger$ correspond to indistinguishable fermions. However, in the SFT they are replaced by commutations because this theory deals with three kinds of distinguishable particles. Therefore the SFT has to be antisymmetrized {\it a posteriori}. 

The effective repulsion in the $U_d=\infty$ limit is a kinetic, time-lag effect \cite{ka1,fr3,bb1} in the sense that one particle has to wait for the other one to leave the Cu site in order to cross it. It is therefore appropriate to use the time-dependent theory. The convenient operative choice is the time-dependent many-body perturbation theory which treats $t_{pd}$ to the infinite order of perturbation. In this theory it is essential \cite{agd} to start from the {\it nondegenerate} unperturbed ground state of the $t_{pd}=0$ Hamiltonian $H_0$. The slave particle representation via slave bosons or slave fermions has to be chosen accordingly and we turn first to this question.

The unperturbed slave particle ground state $|G_0\rangle$ has itself to be metallic and at $t_{pd}=0$ it is the direct product of the states of three kinds of particles, $|G_0^f\rangle\otimes|G_0^b\rangle\otimes|G_0^p\rangle$. The only way to ensure translational and local gauge invariance in the unperturbed state is by taking $n_b^{(0)}=0$. $|G_0^p\rangle$ is then the usual Hartree-Fock (HF) state of $2n_p^{(0)}=1+x$ fermions $p_{\vec k,\sigma}^{(i)\dagger}$ in the cosine band associated with $\varepsilon_p$ and $t_{pp}$, which parametrize $H_{0p}$. This band is filled up to the chemical potential $\mu_{1+x}$ and is usually folded artificially in two $i=l,\tilde l$ oxygen bands $\varepsilon_{p\vec k}^{(i)}$  in the CuO$_2$ Brillouin zone. Such a $|G_0^p\rangle$ is, of course, nondegenerate in both slave particle representations.  The $n_f^{(0)}=1$ $d^{10}$ state on Cu is denoted by $f_{\vec R}^\dagger|\tilde 0\rangle$ , where $|\tilde 0\rangle$ is the auxiliary vacuum on Cu. $|G_0^f\rangle$ is a nondegenerate state of the $t_{pd}=0$ slave particle Hamiltonian $H_{0\lambda}$ only if  $f_{\vec R}$-particles are taken as spinless fermions rather than as spinless bosons. Only then can $|G_0^f\rangle$ be equally simply expressed in terms of Fourier transforms $f_k^\dagger$ of the operators $f_R^\dagger$ since it is only then that we have

\begin{equation}
|G_0^f\rangle=\prod_{\vec R}f_{\vec R}^\dagger|\tilde 0\rangle=\prod_{\vec k}f_{\vec k}^\dagger|\tilde 0\rangle\;,\label{Eq01}
\end{equation}

\noindent up to an unimportant phase factor. Here the product over $\vec k$ covers the whole CuO$_2$ Brillouin zone. In other words, the Mott state of spinless fermions is equivalent to the filled (dispersionless) band of those particles.  When the $f_{\vec k}^\dagger$ are chosen as spinless fermions, $|G_0^b\rangle$ is the state with no $b$-bosons, i.e. $n_b^{(0)}=0$ or, equivalently, the $d^9$ states $b_{\vec R,\sigma}^\dagger|\tilde 0\rangle$ are absent in $|G_0\rangle$. The overall unperturbed $n_b^{(0)}=n_d^{(0)}=0$  translationally and locally gauge invariant ground state  is thus {\it nondegenerate}. The price of this important achievement is, however, that the SFT must be antisymmetrized {\it a posteriori}.

Equation~(\ref{Eq01}) represents the key step of the present approach. Once the unperturbed ground state $|G_0\rangle$ is expressed in the momentum space so too should the translationally and locally gauge invariant slave particle Hamiltonian $H_\lambda=H_0-\lambda N+H_I$, $H_0=H_{0p}+H_{0\lambda}$,

\[H_0=\sum_{i,\vec k,\sigma}\varepsilon_{p\vec k}^{(i)}p^{(i)\dagger}_{\vec k,\sigma} p^{(i)}_{\vec k,\sigma}
+\sum_{\vec k,\sigma}(\varepsilon_d+\lambda)b^\dagger_{\vec k,\sigma} b _{\vec k,\sigma}+\lambda\sum_{\vec k}f^\dagger_{\vec k}f_{\vec k}\]

\begin{eqnarray}
H_I&=&\frac{i}{\sqrt N}\sum_{i,\sigma,\vec k,\vec q}
t^{(i)}_{pd}(\vec k)b_{\vec k+\vec q,\sigma}^\dagger f_{\vec q}
p_{\vec k,\sigma}^{(i)}+\hc\nonumber\\
t_{pd}^{(i)}(\vec k)&=&t_{pd}\sqrt 2
\left(|\sin{\frac{k_x}{2}}|\pm|\sin{\frac{k_y}{2}}|\right)\;.\label{Eq02}
\end{eqnarray}

\noindent where $N$ is the number of the CuO$_2$ unit cells. The action \cite{agd} of $H_I$ on the state of Eq.~(\ref{Eq01}) is then simply defined in the full momentum representation. Provided only that it is convergent, the ensuing time-dependent slave fermion perturbation theory (SFT), in terms of $H_I$, generates then asymptotically the exact SFT ground state, which is translationally, time inversion and \cite{bb2} locally gauge invariant. We recall however that such a procedure replaces the Cu-O anticommutations by commutations.

This omission is, however, irrelevant in the lowest order of the Dyson perturbation theory \cite{agd}. The reason is that the Cu site is initially empty, while one Cu hole is required for the anticommutator (and two for $U_d$ interaction). The $r=1$ $pdp$ or $dpd$ propagators (particles created and annihilated on O or Cu sites respectively) correspond to the lowest order irreducible Dyson $p$-self-energy proportional to the convolution $B_\lambda^{(0)}*F_\lambda^{(0)}$ of the bare slave particle propagators. If the chemical potential of the $p$-fermions $\mu^{(1)}$ is determined subsequently from the relation $n_d^{(1)}+2n_p^{(1)}=1+x$, keeping in mind that ultimately $n_d=n_b$ in the SFT, the $pdp$ and $dpd$ propagators become equivalent to those of the $t_{pd}$ hybridized HF theory. The HF result is characterized by three $i=L,I,U$ branches of poles $\omega_{\vec k}^{(i)}$ of the free three-band \cite{mr1} model, with large Fermi arcs associated with the lowest $L$-band. The corresponding spectral weights $z_{\vec k}^{(i)}$ are given in terms of the departure of $\omega_{\vec k}^{(i)}$, $i=L,I,U$, from the unhybridized $\varepsilon_d$ and $\varepsilon_{\vec k}^{(i)}$, $i=l,\tilde l$. For example, the spectral weight of the $L$-band in the $r=1$ $pdp$ propagator is

\begin{equation}
z_{\vec k}^{(L)}=\frac{(\omega_{\vec k}^{(L)}-\varepsilon_d)^2(\omega_{\vec k}^{(L)}-\varepsilon^{(l)}_{p\vec k})
(\omega_{\vec k}^{(L)}-\varepsilon^{(\tilde l)}_{p\vec k})}
{t^2_{pd}(\omega_{\vec k}^{(I)}- \omega_{\vec k}^{(L)})(\omega_{\vec k}^{(U)}- \omega_{\vec k}^{(L)})}
\;.\label{Eq03}
\end{equation}

\noindent The $t_{pd}^{-2}$ normalization of $z_{\vec k}^{(i)}$ is chosen so that the hopping from the O-sites to the Cu-site is associated with $t_{pd}$, i.e. the  $\vec k$-dependences of $t_{pd}^{(i)}(\vec k)$ are absorbed in $z_{\vec k}^{(i)}$. It is noteworthy that, due to anticrossing \cite{mr1}, the $z_{\vec k}^{(i)}$ are regular functions of $\vec k$. The lowest order SFT generates the HF propagators irrespectively of the average HF occupation $n_d^{(1)}$ of the Cu-site. On the other hand, it will be seen below that $n_d^{(1)}\approx n_f^{(1)}$ only when $n_d^{(1)}$ is small and similarly then $Q^{(1)}=n_f^{(1)}+n_b^{(1)}\approx1$. The SFT thus converges quickly only when $n_d^{(1)}$ is small. It is therefore important to keep in mind that \cite{fr1} $n_d^{(1)}\approx1/2$ at $|x|$ small for $t_{pd}\gg\Delta_{pd}$ and that finite $t_{pp}$ decreases \cite{mr1} it further. The useful overall rule of thumb is that $n_d^{(1)}<1/2$ as long as the HF chemical potential $\mu^{(1)}$ falls below the vH singularity at $\omega_{vH}$ in the lowest $L$-band i.e., as long as $x<x_{vH}$ where \cite{mr1} $x_{vH}\sim-t_{pp}$ is the positive doping required to reach the vH singularity.

%Figure 01; File fig01.eps
\begin{figure}[tb]

\begin{center}{\scalebox{0.31}
{\includegraphics{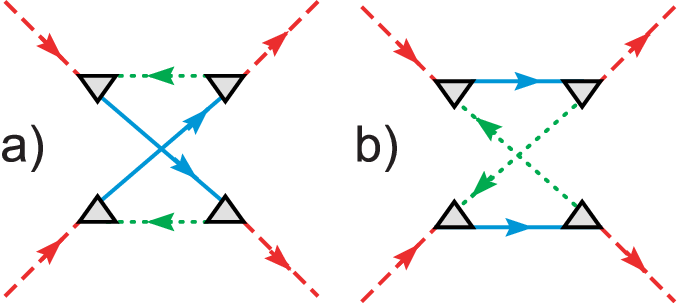}}}
\end{center}

\caption{Effective scattering vertex of two $pdp$ propagators. Triangles denote multiplication by $t_{pd}$. Internal blue and green lines are the dispersionless free propagators of the $b$- and $f$-particles in the state of Eq.~(\ref{Eq01}), respectively, with appropriate arrows of time.\label{Fig01}}

\end{figure}

The central quantity for further expansion is the effective interaction between the $p$-particles shown in Figs.~\ref{Fig01}a,b. It expresses the fact that two holes cannot hop {\it simultaneously} to 
the same Cu-site. When the $t_{pd}$-anticrossing of the $d$-level and the oxygen band strongly affects the occupied states in the $p$-bands, it is appropriate \cite{bb2}, as assumed hereafter, to consider the processes in Fig.~\ref{Fig01} as the interactions between the $r=1$ hybridized $pdp$-propagators normalized according to Eq.~(\ref{Eq03}). The triangular vertices in Figs.~\ref{Fig01} are then to be associated with $t_{pd}$. The internal arrows of time explicitly account for the fact that the bosons can only be created and the spinless fermions only annihilated in the unperturbed ground state in Eq.~(\ref{Eq01}). This defines unambiguously the temporal structure of the internal squares $\Lambda_0^a$ and $\Lambda_0^b$ in Figs~\ref{Fig01}a,b. E.g. $\Lambda_0^b$ of Fig.~\ref{Fig01}b is given by  

\begin{equation}
\Lambda_0^b=\frac{2\varepsilon_d-\omega_1-\omega_2}{\Pi_j(\varepsilon_d-\omega_ j-i\eta)}
\;,\label{Eq04}
\end{equation}

\noindent Here $\omega_j$ denote the external frequencies which run counterclockwise around $\Lambda_0^b$ in Fig.~\ref{Fig01}b with $\omega_1$ in the upper left corner. The result (\ref{Eq04}) is independent of $\lambda$ in Eq.~(\ref{Eq02}). This illustrates one general satisfactory feature \cite{bb2} of the perturbative SFT, namely that it is independent of $\lambda$ order by order even though the local gauge invariance is only asymptotical.

It is important to note, however, that spin is conserved in the triangular vertices of Figs.~\ref{Fig01}a,b i.e., that in the SFT two $pdp$-particles cannot hop simultaneously to the same Cu site through the assistance of slave particles, irrespective of their spin. Contrary to that, for the original Emery Hamiltonian with $U_d$, it is the Pauli principle which forbids two particles in the triplet configuration to occur simultaneously on the Cu-site irrespective of $U_d$. The action of the local two-particle interaction on the fermions of the same spin vanishes thus identically after antisymmetrization \cite{msh}. This discrepancy with the SFT stems clearly from the omission of the Cu-O anticommutations in the SFT and the triplet scattering has to be removed \cite{msh,bb2} from the SFT in order to make it correspond to the original problem. The SFT modified in this way, i.e., antisymmetrized a posteriori, is named the MSFT here.

There are, however, processes, additional to the lowest $r=1$ HF hybridization, for which this Cu-O anticommutation is irrelevant. This is best seen upon closing one $t_{pd}$ $pdp$-hybridized line in each of Figs.~\ref{Fig01}a,b. The effective interaction $\Lambda_0^{a,b}$ is then absorbed in the Dyson bubble renormalization of the slave particle propagators $B_\lambda^{(1)}$ and $F_\lambda^{(1)}$, which is insensitive to the omission of the Cu-O anticommutations. In addition, those propagators are local but each exhibits temporal disorder \cite{bb2}, $b\leftrightarrow f$ symmetrically.  As easily seen,\cite{bb2} the associated $n_b^{(1)}$ is small for $n_d^{(1)}$ small and it follows that $Q^{(1)}\approx1$.  The convolution $B_\lambda^{(1)}*F_\lambda^{(1)}$ is also local and defines the next order ($\lambda$-independent) irreducible $p$-self energy. In the first place, for $n_d^{(1)}$ small this gives the band narrowing $t_{pd}^2\rightarrow t^2_{pd}(1-n_d^{(1)}/2)$ in the $r=2$ $pdp$ and $dpd$ propagations (half of that obtained in the MFSB approaches \cite{ko1,mr1}). In addition the $b$- and $f$-disorders convolute (causally) in frequency. This introduces the imaginary continuum in the $r=2$ $pdp$ and $dpd$ propagations,\cite{bb2} which is causal in time but incoherent in space. In other words, this continuum arises from the spatially incoherent, local, dynamic $d^{10}\leftrightarrow d^9$ charge-transfer fluctuations. The leading disorder contribution falls in the range $2\omega_{\pi,\pi}^{(L)}-\varepsilon_d<\omega<2\mu^{(1)}-\varepsilon_d$, well below the Fermi level. The clearly distinguishable signature of large $U_d$ in the time-dependent perturbation theory is thus the local dynamic $d^{10}\leftrightarrow d^9$ charge-transfer disorder. This can be generalized \cite{bb2} to $r=3$ and further to all processes in which the interactions of Figs.~\ref{Fig01}a,b can be absorbed in the bubble renormalizations of the slave particle propagators, namely to the non crossing approximation \cite{ni1} (NCA). The broad $d^{10}\leftrightarrow d^9$ continuum is expected then to spread \cite{ni1} all over the spectrum. It is interesting to note that a similar broad continuum is obtained in the large $U_d$ local \cite{zl1} dynamic mean-field theory (DMFT).  

%Figure 02; File fig02.eps
\begin{figure}[tb]

\begin{center}{\scalebox{0.31}
{\includegraphics{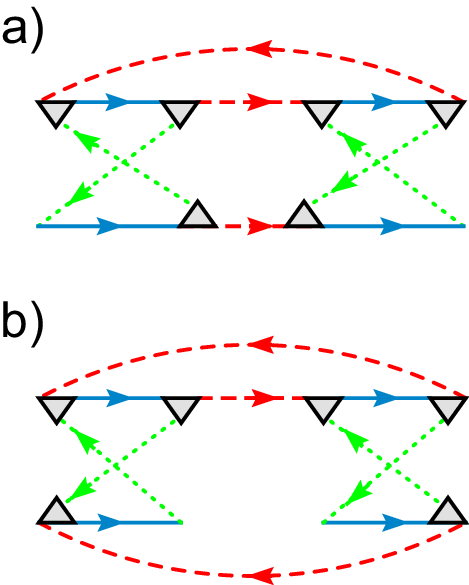}}}
\end{center}

\caption{Contributions to irreducible $p$-self energy of the $r=4$ $pdp$ propagator: (a) in the particle-particle channel; (b) in the particle-hole channel.\label{Fig02}}

\end{figure}

In contrast, the examples of lowest order, irreducible $p$-self energy terms in which the effective interactions of Fig.~\ref{Fig01} cannot be dissolved in the bubble renormalizations of the slave particles are shown in Fig.~\ref{Fig02}. With other $f\leftrightarrow b$ symmetric terms they define the $r=4$ $pdp$ and $dpd$ propagators. The SFT and MSFT start to differ at this level, i.e. the MSFT requires the omission \cite{msh,bb2} of the contributions which involve two $b$-propagators of the same spin within the $\Lambda_0^{a,b}$ squares. In Figs.~\ref{Fig02} this selects out the effect of the {\it singlet} $pdp$-$pdp$ particle-particle and particle-hole correlations on the single particle propagation. Those correlations are best identified on removing the upper receding $r=1$ $pdp$-line (together with its vertices) from Figs.~\ref{Fig02}. What remains are the elementary $pdp$ particle-particle and particle-hole bubbles, which are then subjected to further renormalizations by $t_{pd}^4\Lambda_0^{a,b}$. The MSFT, amounts thus to the summation of a subseries of the locally gauge invariant SFT series, and therefore, is {\it not} manifestly locally gauge invariant. However, if the perturbation procedure for the $B^{(r)}$ and $F^{(r)}$  propagators is carried out $b\rightarrow f$ symmetrically \cite{bb2} the MSFT is at least a conserving \cite{kr1} approximation, $Q^{(r)}=n_b^{(r)}+n_f^{(r)}\approx1$.

The SFT tends to the full and the MSFT at least to the average \cite{kr1} local gauge invariance in the weak coupling manner, provided that $t_{pd}^4(\Lambda_0^a +\Lambda_0^b)$ can be treated as an effective interaction, small with respect to $t_{pd}$ when the external frequencies $\omega_j$ in Eq.~(\ref{Eq04}) fall close to the HF Fermi level $\mu^{(1)}$. Importantly, $n_d^{(1)}$ small makes the effective repulsion $t_{pd}^4\Lambda_0$ small itself. The reason is that $t_{pd}<\varepsilon_d-\mu^{(1)}$ for $\mu^{(1)}<\omega_{vH}$ ($\varepsilon_d-\omega_{vH}$ is linear in $t_{pd}$ for small \cite{fr1} $\Delta_{pd}$ and arbitrary \cite{mr1} $t_{pp}$). According to Eq.~(\ref{Eq04}), the important values of $t_{pd}^4\Lambda_0^{a,b}$ are then less than $t_{pd}$ and, consistently, the internal renormalizations of the slave particle propagators within $\Lambda_0^{a,b}$ can be neglected.

$t_{pd}^4(\Lambda_0^a +\Lambda_0^b)$ is then the effective interaction between the $r=1$ $pdp$ propagators. Such a distribution of the roles of the propagators and the interactions contrasts with the small $U_d$ theory, where the interaction $U_d$ involves the $dpd$ propagators, or with the ansatz  $U_d\rightarrow J_{pd}(\cos{k_x}+\cos{k_y})$ that is often used \cite{sq1}  in the approach from the MFSB side. The $r=1$ $pdp$ propagators with spectral weights of Eq.~(\ref{Eq03}), combined with $t_{pd}^4(\Lambda_0^a +\Lambda_0^b)$, resolve thus the long lasting question \cite{sq1,an2} of what replaces $J_{pd}=4t_{pd}^4/\Delta_{pd}^3$ in the $U_d=\infty$ metallic limit.

Further on, the elementary pdp bubble has the same nonlocal nesting properties(in the denominator)as the elementary $dpd$ bubble but differs from it in the form factor (numerator) \cite{bb1}. The renormalizations of the elementary $pdp$ particle-hole bubble in Fig.~\ref{Fig02}b by $t_{pd}^4\Lambda_0$ are also nonlocal and enhance \cite{bb1} the coherent SDW correlations $\chi_{SDW}$. When the elementary $pdp$ particle-hole bubble in Fig.~\ref{Fig02}b is replaced by such a $\chi_{SDW}$, a structure is obtained which justifies in many respects the semiempirical "single loop" approximation used in the calculation \cite{su2} of the magnetic pseudogap. In particular the couplings of the intermittently added $pdp$ carrier to $\chi_{SDW}$ are thus identified here as $t_{pd}^4\Lambda_0$.

       It can be finally noted that the above choice of the $r=1$ $pdp$ propagators in Fig.~\ref{Fig02} is not unique, i.e. that $r\leq3$ or even the NCA $pdp$ propagators can also be chosen as outsets of the renormalization algorithms according to the value of $x$ under consideration. The NCA emphasizes the $d^{10}\leftrightarrow d^9$ charge-transfer disorder effects while the choices $r=1,2$ favor the coherent SDW correlations.
       
       In conclusion, $t_{pd}^4(\Lambda_0^a +\Lambda_0^b)$ turns out to be the elementary effective local singlet repulsion of two $pdp$-particles. It is the small $\Delta_{pd}$, $n_d^{(0)}=0$ counterpart of the $n_d^{(0)}=1$ superexchange \cite{ge1,za1} $J_{pd}=4t_{pd}^4/\Delta_{pd}^3$ obtained for large $\Delta_{pd}$. The present perturbation theory is flexible enough \cite{bb1,bb2} to deal with the competition between the $d^{10}\leftrightarrow d^9$ disorder and the coherence associated with the incommensurate \cite{bb1} SDW in simple approximations \cite{fr1,bs5,dz2,fr3}. Thus the question left open here, whether or not the omission of the triplet scattering exhausts the a posteriori antisymmetrization of the SFT, i.e. whether the asymptotic behavior of the MSFT is exactly or approximately locally gauge invariant, becomes of little practical importance.

\begin{acknowledgments}

Invaluable discussions with J. Friedel, L. P. Gor'kov, I. Kup\v ci\' c, D. K. Sunko and E. Tuti\v s are gratefully acknowledged.
 
\end{acknowledgments}


\begin{thebibliography}{99}

\bibitem{fr1} J. Friedel, J. Phys. Cond. Matt. {\bf 1}, 7757, (1989).

\bibitem{fr3} J. Friedel and M. Kohmoto, Eur. Phys. J. B {\bf 30}, 427 (2002).

\bibitem{an1} P. W. Anderson, Nature Physics  {\bf 2}, 626 (2006).

\bibitem{em1} V. J. Emery, Phys. Rev. Lett. {\bf 58}, 2794 (1987).
         
\bibitem{ge1} W. Geertsma, Physica B {\bf 164}, 241 (1990); Ph.D. Thesis (University of Groningen, 1979).

\bibitem{za1} F. C. Zhang and T. M. Rice, Phys. Rev. B {\bf 37}, 3759 (1987).

\bibitem{ku1} I. Kup\v ci\' c, S. Bari\v si\' c and E. Tuti\v s, Phys. Rev. B {\bf 57}, 8590 (1998).

\bibitem{go1} L.P. Gor'kov and A. V. Sokol, JETP Lett. {\bf 46}, 420 (1987).

\bibitem{bs5} S. Bari\v si\' c and Batisti\' c,
         Pysica Scripta. {\bf T27}, 78 (1989).
         
\bibitem{mr1} I. Mrkonji\' c and S. Bari\v si\' c, Eur. Phys. J. B {\bf 34}, 69 (2003); {\bf 34}, 441 (2003). 

\bibitem{su2} D. K. Sunko and S. Bari\v si\' c, Phys. Rev. B {\bf 75}, 060506(R) (2007).

\bibitem{bn1} S. E. Barnes, J. Phys. F {\bf 6}, 1375 (1976). 
	
\bibitem{ko1} G. Kotliar {\it et al}, Physica C {\bf 153-155}, 538 (1988).

\bibitem{li1} Y. M. Li, D. N. Sheng, Z. B. Su and L. Yu, Phys. Rev. B  {\bf 45}, 5428 (1992).

\bibitem{bb1} S. Bari\v si\' c and O. S. Bari\v si\' c, Proceedings of ECRYS 2008, Physica B, (2008) in press.

\bibitem{ka1} J. Kanamori, Prog. Theor. Phys. {\bf 30}, 275 (1963).

\bibitem{agd} A. Abrikosov, L.P. Gor'kov, and I.E. Dzyaloshinskii, {\it Methods of Quantum Field Theory} (Dover Publ., Inc., New York, 1963).

\bibitem{bb2} S. Bari\v si\' c and O. S. Bari\v si\' c, unpublished.

\bibitem{msh} A. Messiah, Quantum Mechanics II, ch. XIV (North-Holland Publ. Co. 1967).

\bibitem{ni1} H. Nik\v si\' c, E. Tuti\v s and S. Bari\v si\' c, Physica C {\bf 241}, 247 (1995).

\bibitem{zl1} M. B. Z\" olfl, Th. Maier, Th. Pruchke, and J. Keller, Eur. Phys. B {\bf 13}, 47 (2000).

\bibitem{sq1} Q. Si, Y. Zha, K. Levin, and J. P. Lu, Phys. Rev. B {\bf 47}, 9055 (1993).

\bibitem{an2} P. W. Anderson, Adv. Phys. {\bf 46}, 3 (1997).

\bibitem{dz2} I. E. Dzyaloshinshii {\it et al}, ZhETF {\bf 94}, 344 (1988).
         
\bibitem{kr1} J. Kroha and P. W\" olfle, cond-mat/0410273 (unpublished).

%\bibitem{tu1} E. Tuti\v s, Ph.D. Thesis (University of Zagreb, 1994).







\end{thebibliography}
\end{document}